# Artificial Intelligence in archival and historical scholarship workflow: HTS and ChatGPT

Salvatore Spina
salvatore.spina@unict.it

**Abstract**

This article examines the impact of Artificial Intelligence on the archival heritage digitization processes, specifically regarding the manuscripts' automatic transcription, their correction, and normalization. It highlights how digitality has compelled scholars to redefine Archive and History field and has facilitated the accessibility of analogue sources through digitization and integration into big data. The study focuses on two AI systems, namely Transkribus and ChatGPT, which enable efficient analysis and transcription of digitized sources. The article presents a test of ChatGPT, which was utilized to normalize the text of 366 letters stored in the "Correspondence" section of the Biscari Archive (Catania). Although the AI exhibited some limitations that resulted in inaccuracies, the corrected texts met expectations. Overall, the article concludes that digitization and AI can significantly enhance archival and historical research by allowing the analysis of vast amounts of data and the application of computational linguistic tools.

**Introduction**

As a language –as every system of communication and cultural experience– also Humanities are being drawn into the world of computing, and the digital turn has pushed scholars to redefine Language, Arts, and History disciplines paradigms. Just as words have been assimilated into "second-order abstractions" (*i.e.* numbers) (Cartesio 2003; 1999), the concept of "source" has taken on those forms of communication that characterise the «onlife» (Floridi 2012; 2015) society.

The digital (and digitised) document will be digital historians' *(meta)source* (Fiormonte 2000) to let them write about our present. However, this perspective seems to leave out the analogue one. If we consider archival documents for a moment, we are most inclined to consider them non-existent assets –what is not on the Internet does not exist! They regain their cultural-historical role only when given attention, studied, and analysed by traditional historians.

Nevertheless, the digital turn, digitalisation and digitisation bring them back into the discourse of archivists and historians, activating a path that can place them within the macro-definition of "Big Data" (Graham, Milligan, and Weingart 2015; Frederic Kaplan and di Lenardo 2017; Santoro 2015; Schiuma and Carlucci 2018).

Analogue sources are transformed into images enriched with metadata and digital information, enabling their identification on the web. Their volume is increasingly vast and difficult to manage without the aid of other ICT tools. Two Artificial Intelligences now offer the opportunity to enhance archival and historical workflows. On the one hand, they enable the automatic transcription of large quantities of manuscripts, and on the other hand, they facilitate a more profound analysis that allows for the mining and extraction of extensive information from texts. This enables the reconstruction of past events with an extraordinary wealth of historical details.

Beyond a purely philological-linguistic approach, in the case of historical analysis, source texts require a certain degree of normalization and correction in order to subsequently be utilized in other computational contexts. Unlike textual corpora that pertain to a single subject, such as literary works, archival sources used by historians exhibit a variety of styles and languages, even in the case of written productions attributed to the same individual. For instance, in the case of a monarch whose documentary production is entrusted to secretaries and various individuals, it is not possible to identify a stylistic homogeneity.

In the case, therefore, of epistolography, such as the one under examination, namely the correspondence of the Biscari princes, the use of artificial intelligence could enable scholars to transcribe multiple texts in a short amount of time and obtain an accurate version of them.

Handwritten text recognition systems, such as tranScriptorium (Sánchez et al. 2013), Transkribus (Erwin 2020; Kahle et al. 2017; Milioni 2020; Muehlberger et al. 2019), and eScriptorium (Kiessling et al. 2019)[1], have expanded the prospects for archivists and historians. For the latter, digitalization has undoubtedly facilitated access to numerous archival documents, creating a condition where historical events can be reconstructed with greater detail beyond the possibility of working with transcribed texts. For archivists, artificial intelligence has enabled the development of digitization projects that go beyond the simple photographic representation of documents, as exemplified by the Archive of the City of Amsterdam, the Swedish National Archive, and the Regesta Pomeraniae Monastica.

However, automatic transcription is not the only challenge related to the world of digitization. The ability to rapidly obtain the transcribed (digital) version of numerous documents from archival collections poses a "physiological" limit for researchers in quickly analyzing their content. While not intending to reopen the "quantity-quality" debate here, it is true that digitalization severely tests the Droysenian capacity of historians to connect facts through the use of analogy, imagination, and subjective interpretation (Droysen 1868).

Quickly transcribing, in digital format, numerous documents –albeit with a certain margin of error– generate an uncontrollable mass of data and information that cannot be easily organised and reviewed and does not let identifying any patterns.

Historical research –since Herodotus– is based on a close reading methodology and a limited number of documents to be analysed. A lot of digitization projects –like the Library of Congress's collection of millions of newspaper pages and the Finnish Archives' court records dating back to the 19th century– pushed historians to start using machine learning (deep neural networks in particular) to control, organize and examine historical documents.

So, «Big data of History» –for researchers– is at once a problem and an opportunity: there is much more information, and often there has been no existing way to sift through it. ICT is a great solution to the problem. It is enough to think that historians describe the plague epidemic in the seventeenth century in Venice relying on archival sources that figured out data and information from documents that record only a few days of that terrible experience. During the "Venice Time Machine" project (Frédéric Kaplan 2015), digitisation has instead allowed scholars to trace three years of events and incidents, increasing knowledge about the epidemic (Lazzari et al. 2020).

HTR tools, on the other hand, also have their limitations. The models are constantly trained and enhanced. Their use has undoubtedly had a profound impact on the Humanities research field, in particular, Archival Science (Nockels et al. 2022). However, it is necessary to consider that the documentary material processed by these technologies is not correctly transcribed in its entirety. Each model has an error range (Character Error Rate) that requires scholars to carry out a thorough revision for two reasons: (1) in the case of wanting to disseminate a digital edition of a complex (large or small) documentary or an epistolary corpus, or other materials, through the internet; (2) in the case of a detailed historical or linguistic analysis that requires accurate and normalized texts, regardless of their potential publication and/or dissemination. From a perspective of mass digitization, such correction cannot be entrusted to an individual scholar, as their capacity would be limited to a few documents, thereby prolonging the research timeline.

However, nowadays, Computer Science and ICT development companies may have unconsciously found a solution: we are referring to LLM (Large Language Model), which is an artificial intelligence that, due to its characteristics related to the construction and formulation of texts, could be configured as the best tool for correcting lengthy texts.

OpenAI was founded in 2015. Since then, significant funding has enabled the company to develop InstructGPT (Ouyang et al. 2022), followed by ChatGPT (Alshater 2022; Jiao et al. 2023; Pavlik 2023; Zhai 2022; Sobania et al. 2023), an innovative AI system capable of engaging in dialogue with humans, thus realizing Alan Turing's vision (Turing 1950). From its early stages of development, GPT has emerged as one of the leading Language Models (LLMs). However, it is important to note that regardless of the testing it undergoes, the AI ChatGPT lacks any semantic understanding. Any test involving "reasoning" is inherently flawed *a priori*.

---

[1] Several articles have analyzed the differences between HTR (Handwritten Text Recognition) software and systems, with a particular focus on the two mentioned here, Transkribus and eScriptorium (Huff and Stöbener 2022; Maarand et al. 2022). Additionally, a report of the study day held at the Bibliothèque nationale de France on May 9, 2022, provides an insight into the development of these two platforms (Gautier et al. 2022).

GPT is not a biological entity. Despite its ability to engage in conversation –it almost seems like discussing with another sentient being– it is not an intelligent agent that relies on semantic understanding and logical inferences. ChatGPT is an AI system constructed upon models that enable it to generate syntactically accurate texts and statements (Floridi and Chiriatti 2020). However, like other LLMs, ChatGPT is susceptible to issues of hallucination (Radford et al. 2019; Muennighoff et al. 2022) throughout its outputs.

Therefore, its strength lies in the feature to fix and reformulate a text, or its parts, syntactically correctly (Floridi and Chiriatti 2020).

In this regard, the testing procedure involved requesting the correction of the Biscari epistolary collection, which has been digitized and automatically transcribed using the Transkribus AI. The corpus comprises 366 letters sent by various individuals to the Paternò Castello, princes of Biscari. The letters were automatically transcribed using the "Transkribus Italian Handwriting M1" public model, which exhibits a Character Error Rate (CER) of 12.50% for the "train set" and 6.70% for the "validation set". Subsequently, the transcriptions were downloaded in text format without undergoing any manual corrections or training aimed at reducing the error rate.

The subsequent phase involved initiating a "chat"[2] session to prompt ChatGPT to analyze each letter to assess whether this OpenAI tool can correct and normalize them.

**Test**

The prompt given to ChatGPT was to correct a text without linguistic instructions, historical references, or training on how to provide expansions of abbreviations.

The type of analysis required does not deviate from the capabilities of AI, and the obtained corrections, which meet expectations, allow us to assert, without hesitation, that the LLM ChatGPT can be a useful tool for correcting historical texts, especially those transcribed automatically. However, a problem arises regarding the length of the texts that require correction, which contradicts one of the fundamental principles of the digital era: the ability to correct, analyze, and process vast amounts of data, be it text or numbers.

At the beginning of the test, we use GPT-3, which could only analyze and correct texts of up to 1000 characters. In our case, this meant processing one letter at a time. This limitation, on the one hand, prevents historians from obtaining a correction for the entire body of texts in a single phase. On the other hand, it does not allow a revision process that takes into account linguistic and syntactic correlations or any relevant data within the corpus that could contribute to a more thorough correction. Additionally, if scholars wanted to correct multiple texts, they would have to provide as many prompts as the number of letters required.

The latest update of the AI (May 2023), as stated by OpenAI, no longer has this limitation (Figures 1, 2, and 3). The company claims that «GPT-4 is capable of handling over 25,000 words of text, enabling use cases such as creating long-form content, engaging in extended conversations, and conducting document research and analysis». This should allow us to correct the entire corpus of letters, which consists of approximately 900,000 characters.

However, once again, the desired outcome cannot be achieved: the character limit present in GPT-3 is still active in subsequent versions (such as GPT-4), and although there has been a significant improvement in computational instructions and text analysis capabilities, the epistology, which comprises 10,662 words, cannot be corrected.

---

[2] Despite its name, ChatGPT has not been primarily used as a chatbot.

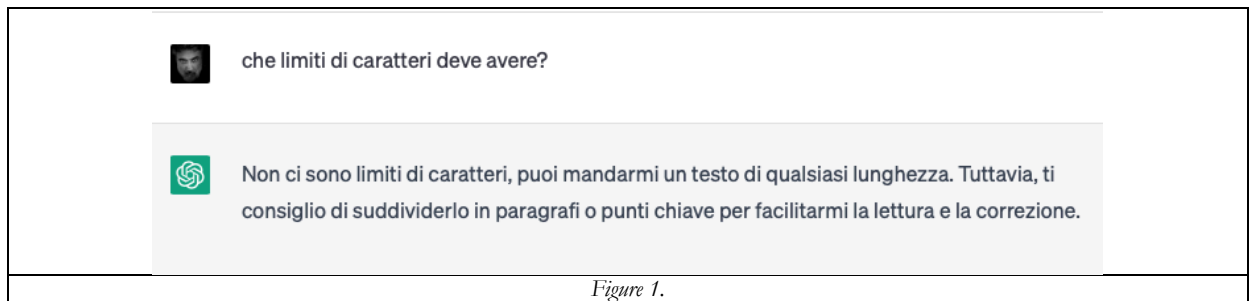

*Figure 1.*

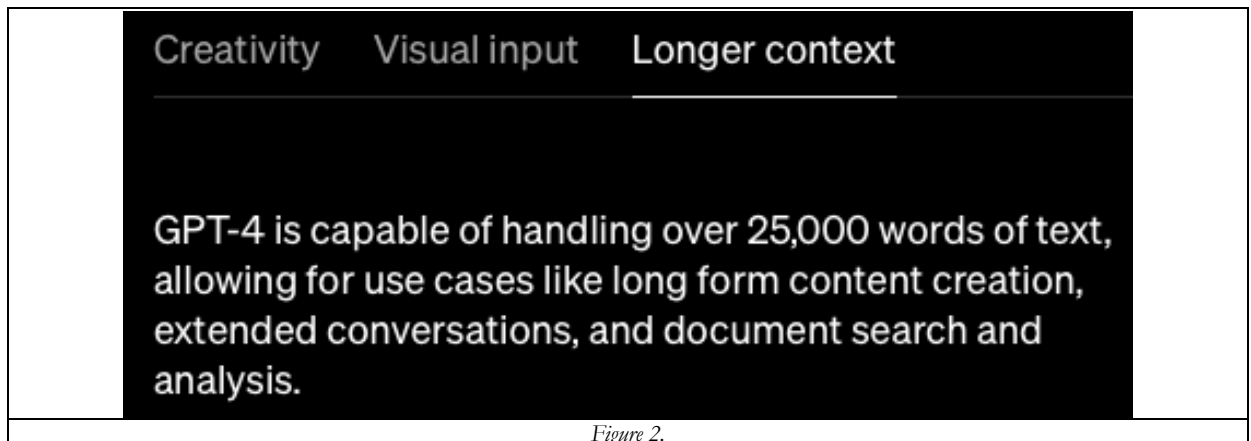

*Figure 2.*

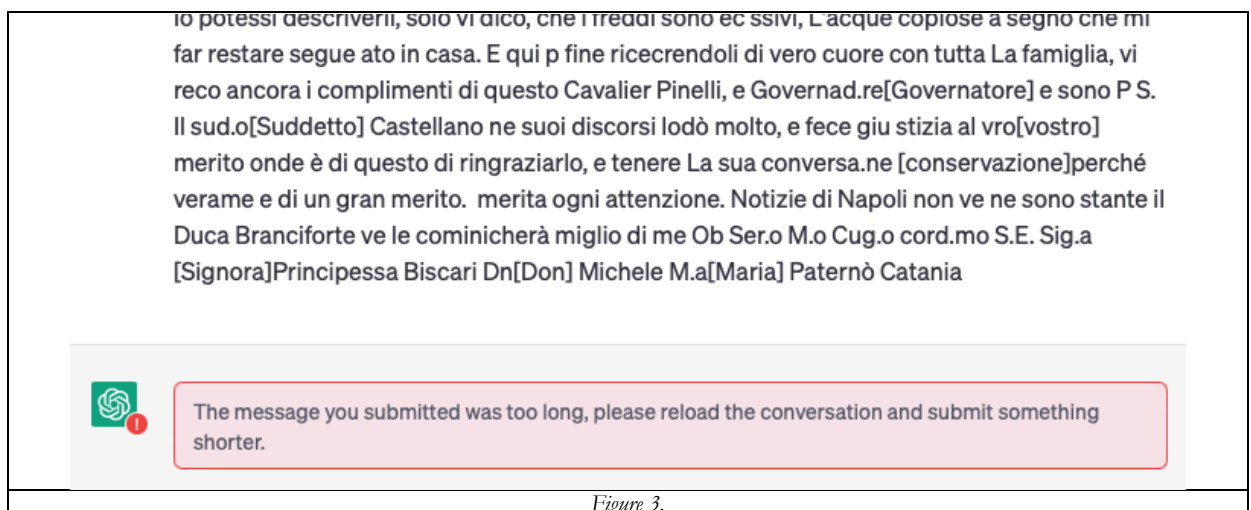

*Figure 3.*

Although ChatGPT is a powerful language generation model, it has a limitation: it lacks access to external information and the ability to browse the internet. Consequently, it cannot provide accurate or up-to-date information on a wide range of topics, and it may struggle to generate responses to complex or unconventional questions (Deng and Lin 2022).

This constraint prevents scholars from analyzing texts that contain information that could enrich the understanding of historical events, as in the case of the letters sent by Michele Maria Paternò to Anna Morso, wife of Ignazio, the 5th prince of Biscari. These letters cover a period (1781-1786) marked by various events related to the history of Sicily, Naples, and the Bourbon court (intrigues, attempts at reforms, management of seismic emergencies such as the reconstruction of the Catania Cathedral, and aspects of religious life), but they cannot offer valuable insights into the history of Sicily in the modern era (Figure 4).

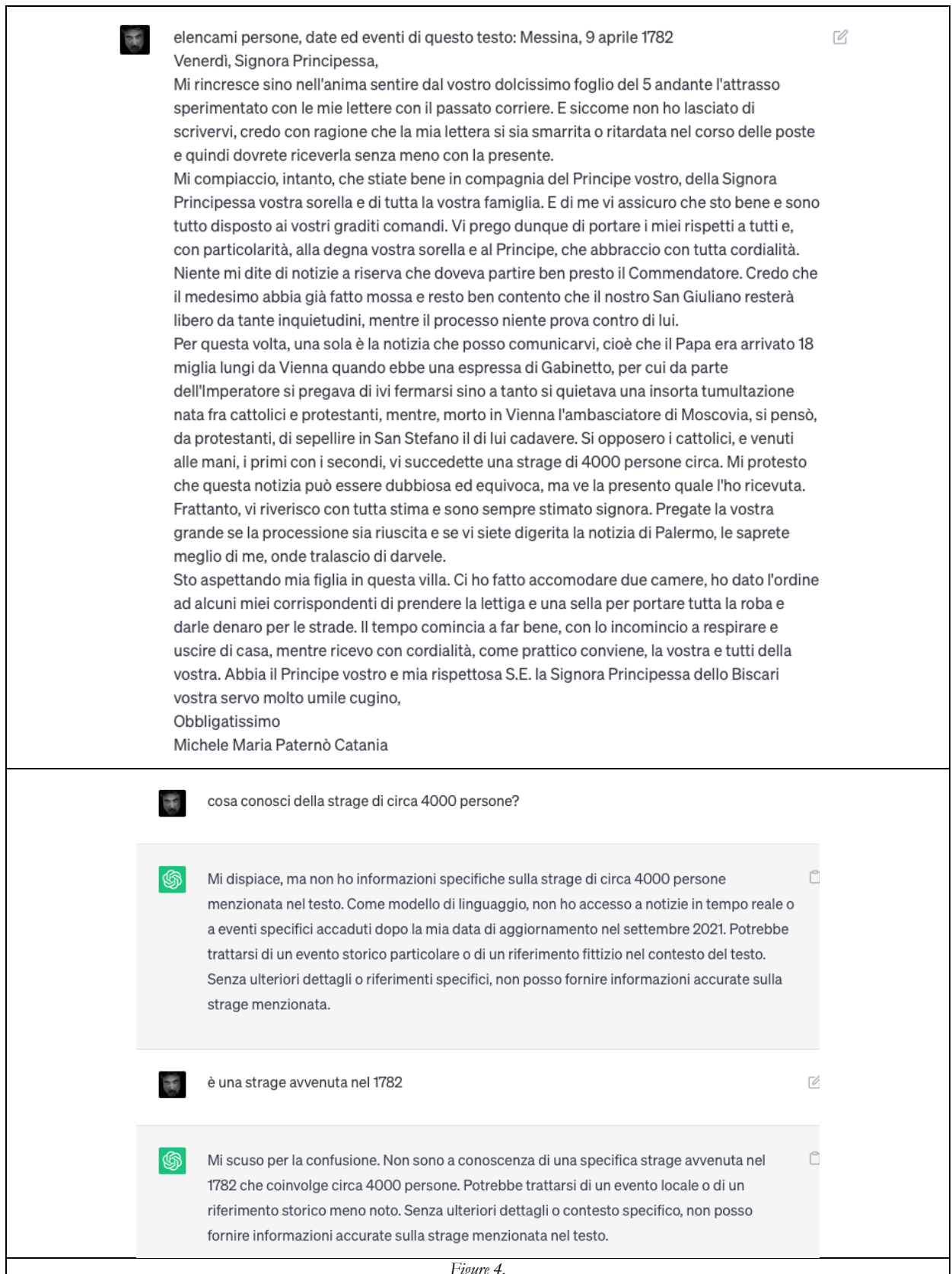

*Figure 4.*

AI cannot perform comparisons between the text and the Web, as Transkribus does. However, it can only operate on the data present within the text being corrected, thus preventing the evaluation of OpenAI's technology in terms of "capability" and "alignment".

The correction, therefore, is necessarily performed on the individual epistles.

ChatGPT first aims for a probability distribution over word sequences or token sequences, which enables it to predict the next word in a sequence. In practical applications, however, there is a clear disconnect between the training of language models (LLMs) and their intended use for valuable cognitive work. While LLMs can generate text, this does not directly correspond to their ability to perform accurate corrections. The text generated by the AI may appear to align with the original text, satisfying researchers' expectations to some extent (Figure 4). The statistical architecture of the machine allows it to generate text where the distribution of tokens conveys the linguistic sense of the text to be corrected. However, without specific instructions from the operator and an unspecified set of instructions specific to the use of a potential "ground truth", in many cases, even if the AI manages to generate text with minor inaccuracies corrected, the "next-token-prediction" may yield incorrect names, inaccurately expand abbreviations, although it can generate the name of the month composed of numbers and letters (8bre, 9mbre, Xmbre) in the body of the letters. Furthermore, due to its inherent function, the AI may introduce names that are not present in the original text or paragraphs that do not faithfully adhere to the original (as shown in Figures 5 and 6).

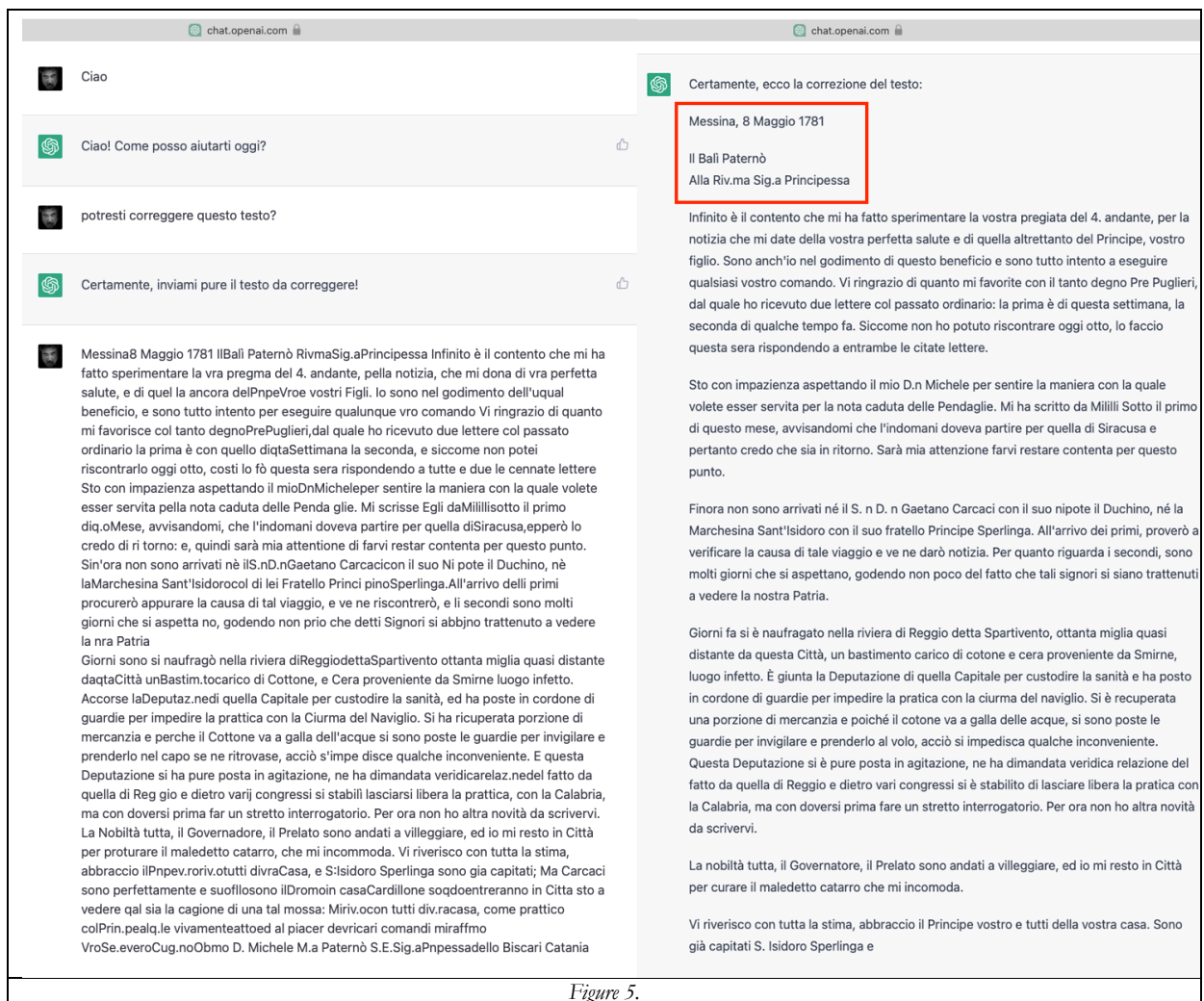

*Figure 5.*

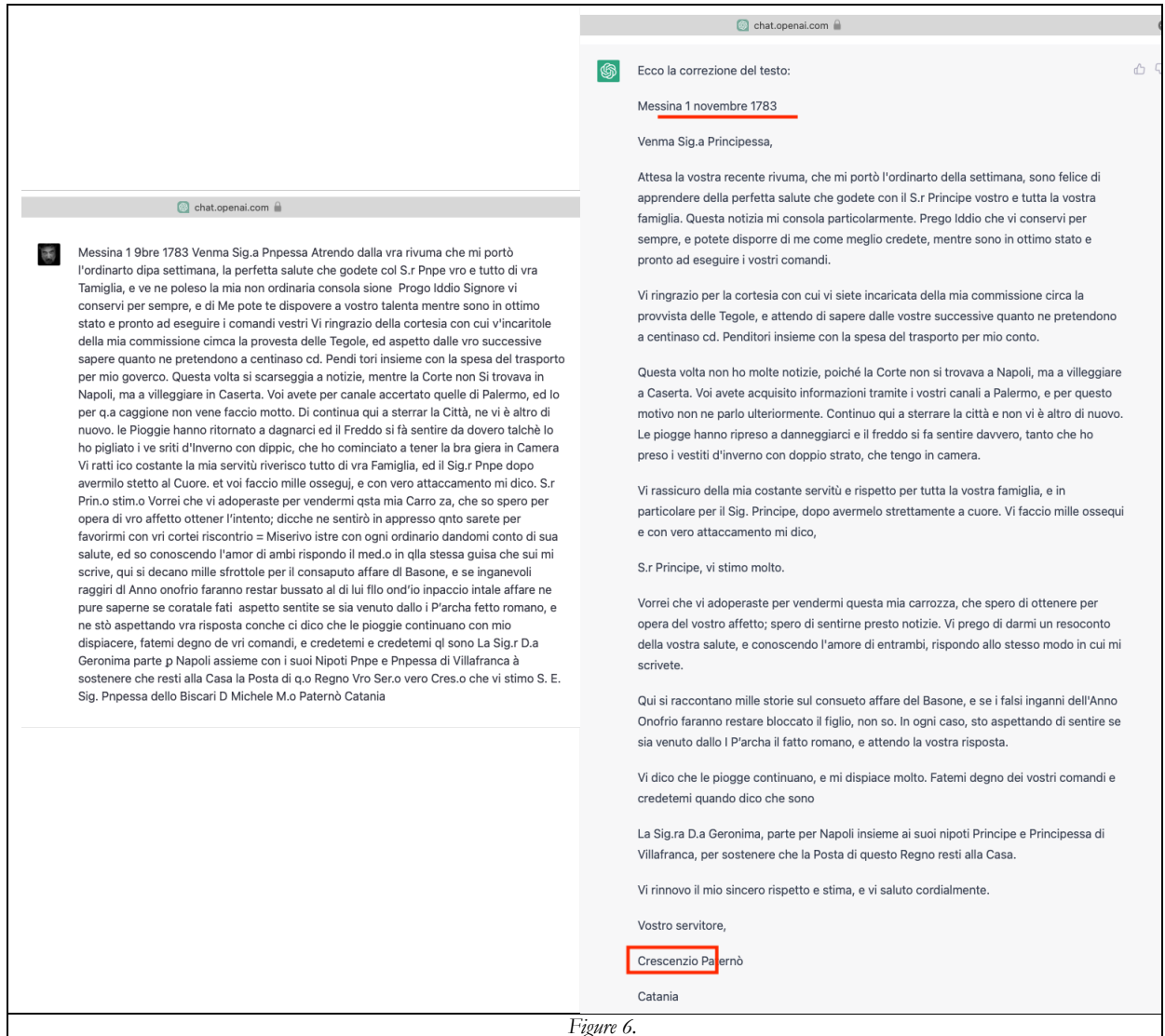

*Figure 6.*

We proceed, therefore, by providing instructions to improve the result, prompting to expand the abbreviations found in the text and correcting those that have been expanded incorrectly (Reinforcement Learning with Human Feedback – RLHF). Subsequently, operators' instructions have successfully enabled the correct expansion of abbreviations in other epistles, resulting in a normalized text of the latter and a list of all abbreviations present in the corpus (Figure 7).

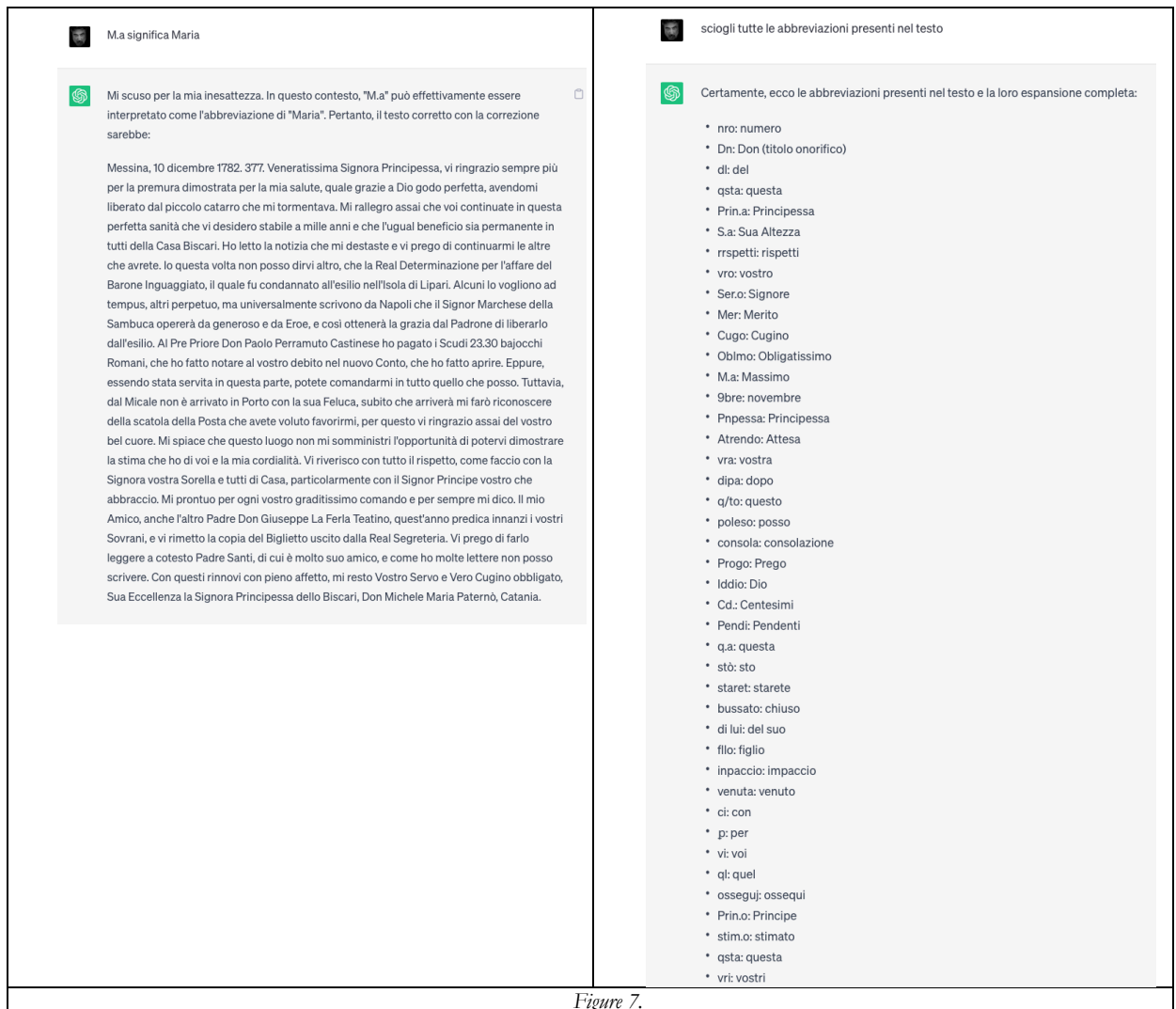

*Figure 7.*

AI enhances numerous fields within the Humanities. While historians can now have the opportunity to work with much larger volumes of data compared to the past, it is even more true that archivists can utilize these computational technologies to improve their workflow. Nowadays, Archival Science is also a "Networked Science". Archival heritage is constantly being digitized and disseminated on the web through websites and digital databases. However, the process of digitization requires a certain level of dynamism based on the principle of constructing a machine-readable version of the documents. But before reaching the completeness of archival information and achieving meticulous and analytical digitization, archivists have the task of creating cards in which, if it is not possible to include the complete transcription of the photographed document, at least useful data can be provided for online research.

The creation of these cards involves identifying specific information, such as in the case of epistolography, the individuals involved (sender, recipient, and other subjects), places (origin, destination, and other locations), dates, and, when possible, events. These pieces of information make the digital archive more dynamic. The work of archivists is therefore meticulous and requires a deep understanding of the written document. However, in this specific case, ChatGPT is a technology that can truly serve humanists.

Although recent studies have highlighted inaccuracies in the recognition of entities in historical documents (González-Gallardo et al. 2023) –possibly because ChatGPT does not have access to information available on the web–, in our case, the AI was able to identify entities (named entity recognition and classification - NERC) with absolute precision, providing details that have guided us in creating detailed records of the documents. In our case, with the assistance of ChatGPT, we have successfully identified the relevant information for compiling records on the "Biscari Epistolography" website (Figure 8).

*Figure 8.*

**Consideration**

Historical scholarship is primarily based on text-based analysis –where the original wording of historical sources often provides new insights and inferences of specific meanings–, while Archival Science is called to a reassessment of the principle of dissemination of the documentary heritage. Thus, both historians and archivists, albeit from different points of view, need to work on correctly transcribed texts. This happens because, the availability of computing normalised-text corpora in modern language offers certain advantages. Firstly, it enables the dissemination of results without the need for manual normalisation. Secondly, it allows for the application of computational linguistic analysis tools that can uncover additional data crucial for accurate storytelling. Tools like Keyphrase Digger are more effective in processing modern-language normalised texts compared to those written in an 18th-century style (as we can see in Figure 9).

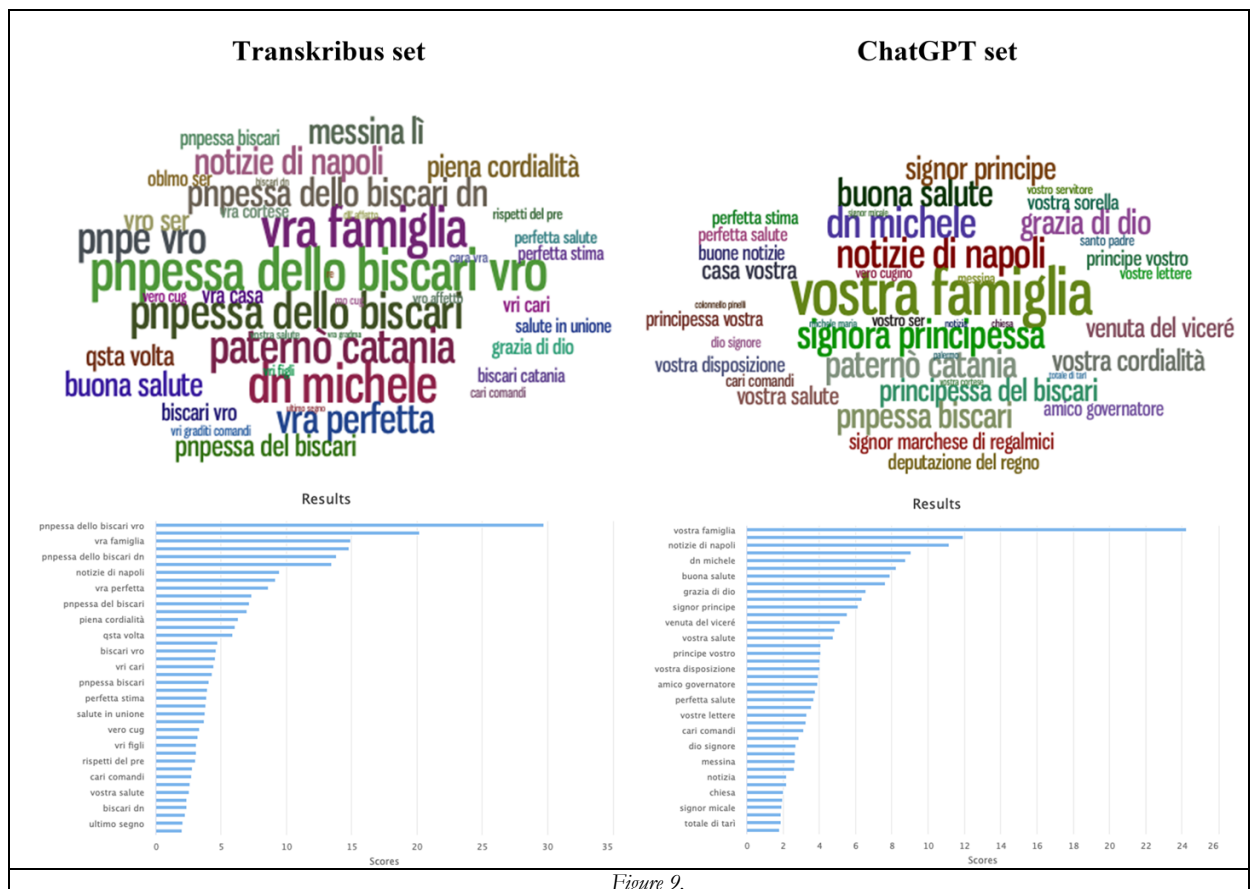
*Figure 9.*

The unique features of 18th-century texts, such as abbreviations, syntactic errors, and incorrect word forms, would require specific modelling to train artificial intelligence. It should be noted that querying neural networks and the semantic web with an 18th-century text might lead to its exclusion.

Artificial intelligence (AI) is a rapidly advancing field within Computer Science, focused on creating intelligent machines capable of human-like thinking and action. While AI has found applications in various domains, History remains a discipline rooted in archival sources. Accurate digitization and automatic transcription projects are essential for providing the necessary information to develop models that effectively assist historians in their research.

However, certain reflections are warranted. The *digital turn* and digitization projects have inundated historians with a vast amount of historical data extracted from digitized sources, posing challenges in managing and analyzing such Big Data from the Past. ChatGPT, despite lacking sentient intelligence, utilizes semantic processing and syntactic rules to generate dialogue that may give the impression of understanding. Although it lacks a biological brain structure, the training on syntactic structures and instructions make GPT-3 a valuable tool for normalizing historical source corpora. Corrections made by the AI, even if unrelated to the text itself, such as reformulating propositions or merging names and abbreviations, do not distort the source but rather unearth new data for historians to explore.

Historical research relies on archival sources, which transmit more than just the linguistic structure of the text; they possess semantic structures. Nevertheless, this does not imply that historical methodology should forego the advantages offered by ITC innovations that ensure accuracy at specific stages of historians' workflow. Transkribus, for instance, is an AI-powered tool used to transcribe digitized archival documents with impressive speed, generating machine-readable files that enable historians to analyze, process, and create digital editions. While ChatGPT will not write "the" History itself, its training equips it to write about History, albeit not at the same level as scholars. It can serve as a valuable tool in the hands of archivists and historians, offering an objective and compelling narrative by normalizing information from digitized, encoded, and automatically transcribed sources.